**Title**: Assessing the intelligibility of vocoded speech using a remote testing framework

**Authors**: Kevin M. Chu, Leslie M. Collins, Boyla O. Mainsah

**Affiliation**: Department of Electrical and Computer Engineering, Duke University, 101 Science Drive, Durham, North Carolina, United States, 27708

**Email Addresses**: kevin.m.chu@duke.edu, leslie.collins@duke.edu, boyla.mainsah@duke.edu

**Abstract**

Over the past year, remote speech intelligibility testing has become a popular and necessary alternative to traditional in-person experiments due to the need for physical distancing during the COVID-19 pandemic. A remote framework was developed for conducting speech intelligibility tests with normal hearing listeners. In this study, subjects used their personal computers to complete sentence recognition tasks in anechoic and reverberant listening environments. The results obtained using this remote framework were compared with previously collected in-lab results, and showed higher levels of speech intelligibility among remote study participants than subjects who completed the test in the laboratory.

## 1. Introduction

Speech intelligibility tests have been traditionally conducted in a controlled laboratory environment. In-lab experiments provide the opportunity to test subjects' speech recognition in a soundproof booth, which minimizes the effects of environmental noise and visual distractions. Additionally, in-lab tests allow the experimenter to use the same computer hardware and same set of headphones across subjects, which allows for precise control over the sound level and frequency response of the stimulus (Stecker et al., 2020). Over the past few years, interest has grown in remote speech intelligibility testing, which allows subjects to participate at more convenient times and circumvents the need to travel to the laboratory (Merchant et al., 2021; Stecker et al., 2020). More recently, the COVID-19 pandemic has made human-based research activities challenging as many in-person studies were suspended to prevent further spread of the virus (Stecker et al., 2020). Thus, it has become more essential than ever to develop a viable remote framework for conducting speech intelligibility tests.

In 2020, the Acoustical Society of America organized the Task Force on Remote Testing for identifying the best practices for conducting human-based acoustics research in a remote setting. The task force created a Wiki that describes issues related to remote testing platforms, stimulus presentation and data collection, and regulatory requirements. One of the major considerations is to identify the hardware and software that will be used for remote data collection. In general, selection of the appropriate hardware and software involves a trade-off between experimental control and convenience. For hardware, one approach is for the experimenter to deliver a specific set of hardware (e.g. headphones and computer) to the subject so that they can complete the listening test at home. Using a specific set of hardware allows the experimenter to use pre-installed software and to precisely control stimulus presentation.

However, this approach is time-consuming because it requires the experimenter to arrange specific times with the subject to drop off and pick up hardware. An alternative approach is to use the subject's personal hardware. While this approach offers less control than using a specific set of hardware, it is more convenient because it does not require the experimenter to deliver equipment (Stecker et al., 2020).

In terms of software, the options range from preconfigured software packages to customized scripts. Preconfigured software packages provide a convenient way for the experimenter to perform standard psychoacoustical procedures (Stecker et al., 2020). For example, the Portable Automated Rapid Testing (PART) is an application that allows the subject to self-administer a variety of psychoacoustical tasks on a portable device (Gallun et al., 2018; Lelo de Larrea-Mancera et al., 2020). These psychoacoustical tasks include but are not limited to speech in quiet, spatial release from masking, and spectro-temporal modulation sensitivity (Gallun et al., 2018). However, the main disadvantage of these preconfigured software packages is that they do not provide the ability to implement customized procedures. In contrast, customized scripts allow the experimenter to conduct experimental procedures that are not available in preconfigured software packages (Stecker et al., 2020). For example, a recent study used a compiled MATLAB script that allows subjects to self-administer a three-alternative forced-choice word recognition task (Merchant et al., 2021). The goal of this study was to measure the binaural intelligibility level difference in normal hearing listeners, which quantifies the improvement in the speech reception threshold under a diotic masker when target speech is presented 180 degrees out of phase, rather than in phase, to both ears (Merchant et al., 2021). However, this study represents just one remote framework, and scripts must be tailored to perform the desired psychoacoustical task.

Listening tests that evaluate the performance of speech processing and speech enhancement algorithms must also transition to a remote framework in order for research efforts to continue during the pandemic. Remotely assessing speech processing algorithms in CI users is challenging because CI users are especially sensitive to reverberation and noise (Hazrati & Loizou, 2012), which are more prevalent in a remote setting (Hughes et al., 2012). This factor is further compounded by the large variability in speech intelligibility measures in CI users that exists as a result of varying degrees of nerve survival, electrode insertion depth, and bone growth, among other factors (Loizou, 2006). In contrast, vocoder simulations in normal hearing listeners allow the experimenter to test CI speech processing strategies without the confounding effects of large inter-subject variability between CI users (Loizou, 2006). To the best of our knowledge, no at-home framework has been validated for normal hearing listeners performing sentence recognition tasks with vocoded speech.

We developed a remote study protocol that includes electronic consent, a standalone application to perform listening tests, and subject payments. We developed a customized compiled MATLAB script that allows normal hearing subjects to complete listening tests remotely using their own computers. The application supports open set sentence recognition tasks that present the spoken sentence and collect the user's free form response. During the study session, we met virtually with subjects to guide them through program installation and the actual listening test. To determine the feasibility of the remote testing framework, we compared the results of the remote study with those obtained during in-person experiments pre-pandemic.

## 2. Methods

*2.1 Subjects*

Twenty-one subjects aged 18-53 (median = 22.0 years) participated in the remote study. This study was approved by Duke University's Institutional Review Board. Subjects were required to be native speakers of American English and to have self-reported normal hearing. Research Electronic Data Capture (REDCap) (Harris et al., 2009), a secure Health Insurance Portability and Accountability Act (HIPAA)-compliant web-based application, was used to administer electronic consent forms and collect demographic information. Subjects were financially compensated for their time via physical checks or electronic gift cards.

*2.2 Signal Processing*

In this study, subjects were presented with either anechoic or reverberant speech stimuli. To generate the reverberant stimuli, anechoic speech was convolved with recorded room impulse responses (RIRs) from the Aachen Impulse Response database (Jeub et al., 2009). We selected RIRs from the office room, lecture hall, and the stairway. For the stairway condition, RIRs were selected at an azimuth of 90 degrees, where the source and receiver are directly facing each other. Table 1 lists the RIR characteristics including room dimensions, source-receiver distances, dummy head recording channel (i.e. left ear versus right ear), reverberation times ($RT_{60}$), and direct-to-reverberant ratios (DRRs). The DRRs were calculated by computing the ratio between the total energy of the direct path component of the RIR (defined as the initial portion of the RIR up through 8ms following the largest amplitude filter coefficient) to the energy of the reverberant reflections (Naylor et al., 2010). The $RT_{60}$s were calculated based on the Schroeder method (Schroeder, 1965) using code provided by (Jeub, 2010). Because reverberant speech is more challenging to understand than anechoic speech, we believe a reverberant condition will better reveal the differences between a remote study and an in-person study.

Table 1: This table shows characteristics of the room impulse responses (RIRs) used in this study.

| Room | Dimensions (L x W x H) (m) | Source-Receiver Distance (m) | Channel | $RT_{60}$ (s) | DRR (dB) |
|---|---|---|---|---|---|
| Office | 5.0 x 6.4 x 2.9 | 3.0 | Left | 0.6 | 0.2 |
|  |  |  | Right | 0.6 | 0.4 |
| Lecture | 10.80 x 10.90 x 3.15 | 5.56 | Left | 0.8 | 0.1 |
|  |  |  | Right | 0.9 | -0.1 |
| Stairway | 7.0 x 5.2 | 3.0 | Left | 1.0 | 1.9 |
|  |  |  | Right | 0.9 | 1.6 |

We simulated CI processing using the Advanced Combination Encoder (ACE) strategy (Vandali et al., 2000) based on a 22-channel Nucleus device using the Nucleus MATLAB Toolbox (Swanson, 2006). The time domain speech signal was partitioned into 8ms frames with a 2ms shift between consecutive frames. Each frame was passed through a short-time Fourier transform and the symmetric frequency bins were discarded, resulting in a 65-dimensional complex spectrum. The complex spectrum was converted into a magnitude spectrum and multiplied by a set of weighted ACE filters to map the 65-dimensional spectrum to a set of 22 frequency bands representing the electrode channels. For each 2ms stimulation cycle, we selected the 6, 8, 9, 10, 11, or 12 channels with the highest energy. These values aimed to simulate typical settings used in the ACE processing strategy (Vandali et al., 2000) and to create more difficult listening conditions to uncover the differences between remote and in-person studies. Time-frequency bins below the psychophysical threshold were zeroed out, while the bins above the comfort level were clipped. Threshold and comfort levels were selected using the default values from the ACE processing map in the Nucleus MATLAB Toolbox (Swanson, 2006). The signal was vocoded using a sine wave vocoder to simulate the spectral resolution available to a CI user. The vocoded speech stimuli were equalized to the same root-mean-square (RMS) value.

*2.3 Remote Listening Tests*

During the study session, subjects met remotely with the experimenter via Duke University's instances of Webex or Zoom. Subjects completed the listening test on their personal computers by installing and running our listening test interface. The testing interface is identical to the one we use for in-lab experiments, but it is packaged as a standalone application for compatibility with subject's personal computers. We used MATLAB Compiler™ (*MATLAB Compiler*, n.d.) to create a standalone executable that allows for the execution of a MATLAB® (*MATLAB*, n.d.) generated program without requiring a full MATLAB® installation. Applications created via MATLAB Compiler™ require MATLAB® Runtime (*MATLAB*, n.d.), which can be automatically downloaded and installer by the installer. The test interface allows the user to load in a file containing vocoded speech stimuli, which must be pre-generated using a MATLAB® script. The interface presents subjects with a series of sentences one-by-one and allows them to type in a free form response. The interface records and saves the responses for offline analysis. Our executable file is compatible with Windows 10 and macOS 10.13.6 and higher.

Subjects were provided with the executable file as well as a separate file containing the speech stimuli. Files were shared via Duke Box, which allows for secure file transfer. The experimenter guided the subject through the installation process and the listening test procedure. Subjects were instructed to share their computer screen during the listening test to ensure that they were on task and to help them debug technical issues. Thus, the remote study framework mimicked an in-person study session where the experimenter sits alongside the participant.

Subjects listened to the speech stimuli either through their own headphones (in-ear or over-ear) or through their computer speakers only if they did not own headphones compatible with our program. Personal hardware was allowed as exact stimulus presentation levels and frequency responses were not required. Subjects were instructed to type as many words as they could understand, and they were provided with feedback after entering their response. To prevent random guessing, subjects were instructed to type "I don't know" for the unintelligible sentences. The study session was divided into a training phase and a testing phase. During the training phase, subjects were first trained to understand vocoded anechoic speech using sentences from the City University of New York (CUNY) database (Boothroyd et al., 1985). During the training phase, speech stimuli were calibrated via psychophysical means by allowing subjects to adjust the volume to a comfortable level, which they were instructed to maintain for the remainder of the study. Training ended after 20 sentences were presented or when the average percent of correct phonemes over three consecutive non-overlapping blocks of five sentences did not improve by more than 10 percentage points, whichever came last. During the testing phase, subjects were presented with sentences from the Hearing in Noise Test (HINT) database (Nilsson et al., 1994). The testing conditions consisted of anechoic and reverberant speech vocoded using different numbers of selected channels as described previously. The order of the conditions and the assignment of HINT sentence lists to conditions were randomized.

At the end of the study session, subjects uploaded their experimental data via Duke Box. The experimenter then guided the subjects through the process to uninstall the listening test interface from their computers.

*2.4 In-Person Listening Tests*

Results of the remote listening tests were compared with our previously unpublished results from in-person listening tests conducted in the laboratory. Nine subjects aged 18-30 (median = 23.0 years) participated in the in-person study. During the in-person tests, subjects were seated in a double-walled soundproof booth and were presented with sentences through the same pair of headphones. The speech stimuli in the training and testing phase were calibrated to 65 dB sound pressure level. As with the remote study, subjects listened to CUNY sentences (Boothroyd et al., 1985) during the training phase and HINT sentences (Nilsson et al., 1994) during the testing phase. Subjects in the in-person study completed only a subset of the listening test conditions that were used in the remote study, which included an anechoic environment and a reverberant office with 8, 10, or 12 selected channels.

*2.5 Data Analysis*

Speech intelligibility was quantified as the percent of correctly identified phonemes. Speech intelligibility scores were mapped to rationalized arcsine units (RAUs) (Studebaker, 1985) prior to running the inferential statistics. The first analysis focused on the results from the remote study, which were analyzed using a two-way repeated measures analysis of variance (ANOVA) with within-subjects factors of the acoustic environment and number of selected channels as well as their two-way interaction. The second analysis compared the results from the remote and in-person studies using a mixed ANOVA with a between-subjects factor of study location, within-subjects factors of acoustic environment and number of selected channels, and two- and three-way interactions. *Post-hoc* analyses were performed using estimated marginal means with Tukey's multiple comparisons correction. The significance level for all statistical tests was set to 0.05.

## 3. Results

Fig. 1 shows the speech intelligibility scores for subjects who participated in the remote study. The main effects of the acoustic environment ($F(1.8, 35.8) = 97.0$, $p < 0.001$, $\eta_g^2 = 0.415$, Greenhouse-Geisser correction) and the number of selected channels ($F(5, 100) = 9.5$, $p < 0.001$, $\eta_g^2 = 0.048$) were statistically significant, but their two-way interaction was not ($F(6.8, 135.8) = 1.5$, $p = 0.174$, $\eta_g^2 = 0.026$). Speech intelligibility was significantly higher in the anechoic environment than all three reverberant conditions ($p < 0.001$). Additionally, speech intelligibility was significantly higher in the office than in the lecture hall ($p = 0.006$) and the stairway ($p < 0.001$). Speech intelligibility was not significantly different between the lecture hall and the stairway ($p = 0.615$). In terms of the number of selected channels, speech intelligibility was significantly lower for 6 selected channels than for 8 to 12 selected channels ($p < 0.004$), whereas speech intelligibility did not significantly differ between 8 to 12 selected channels ($p > 0.05$).

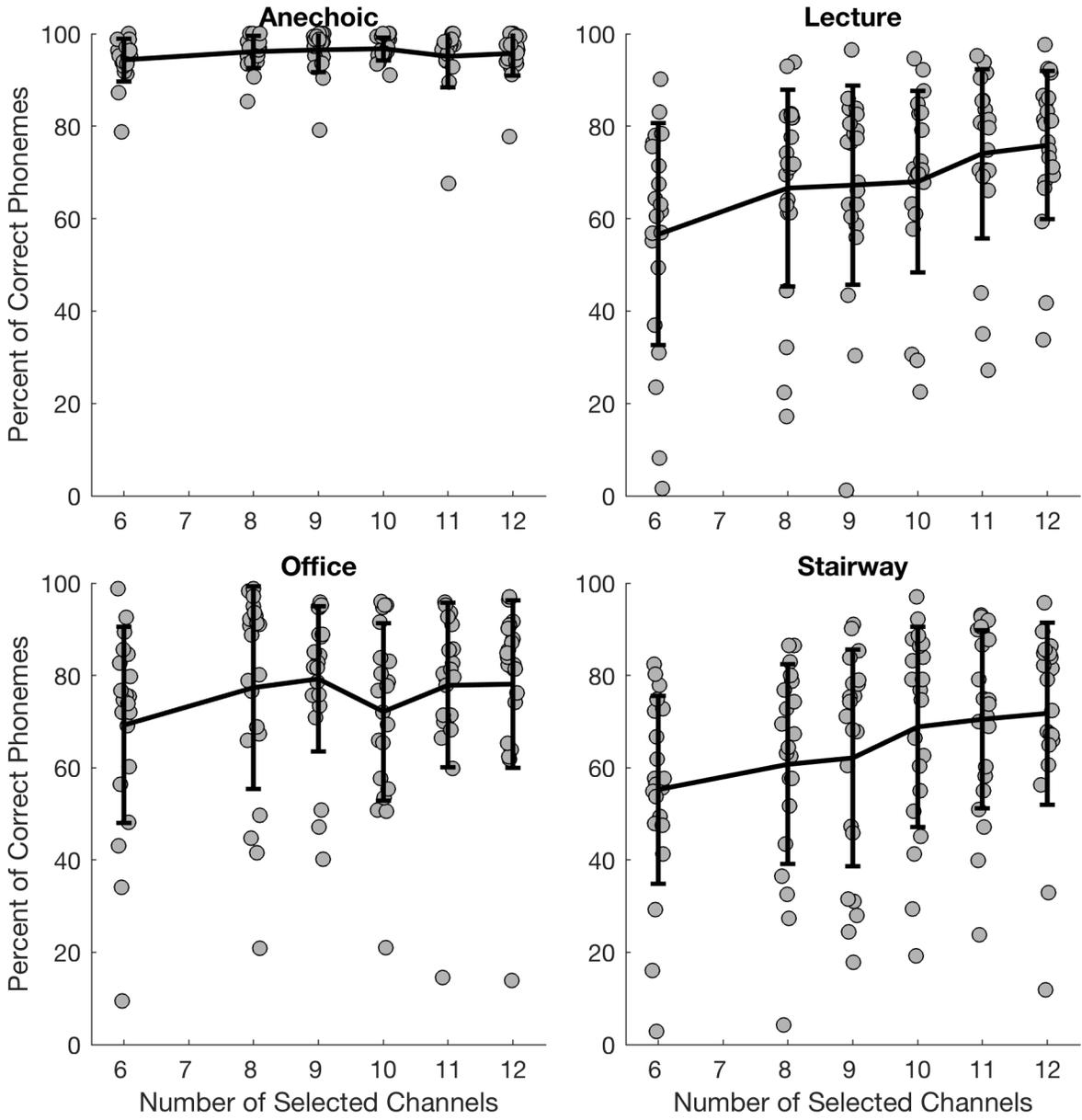

Fig. 1: This figure shows the percent of correctly identified phonemes for all subjects who participated in the remote study. The subplots show subject performance in an anechoic listening environment and three reverberant environments. The solid line shows the mean percent of correctly identified phonemes, error bars show ± 1 standard deviation, and gray circles indicate individual subject performance.

Fig. 2 compares speech intelligibility scores for in-person and remote experiments in an anechoic environment as well as an office. The main effect of room was statistically significant ($F(1, 28) = 126.6$, $p < 0.001$, $\eta_g^2 = 0.558$), but the effects of the number of selected channels ($F(2, 56) = 0.9$, $p = 0.399$, $\eta_g^2 = 0.005$) and the study location ($F(1, 28) = 3.0$, $p = 0.093$, $\eta_g^2 = 0.037$) were not. Despite the insignificant main effect of study location, the two-way interaction between location and room was statistically significant ($F(1, 28) = 6.5$, $p = 0.016$, $\eta_g^2 = 0.061$) as was the interaction between location and the number of selected channels ($F(2, 56) = 3.5$, $p = 0.037$, $\eta_g^2 = 0.019$). In the anechoic environment, speech intelligibility ranged from $95.4 \pm 4.7\%$ to $98.2 \pm 1.6\%$ for the in-person study and $95.8 \pm 4.8\%$ to $96.8 \pm 2.5\%$ for the remote study, and this difference was not statistically significant ($p = 0.703$). In the office, speech intelligibility ranged from $56.3 \pm 12.8\%$ to $68.8 \pm 16.7\%$ for the in-person study and $72.1 \pm 19.2\%$ to $78.1 \pm 18.1\%$ for the remote study, and this difference was statistically significant ($p = 0.004$). All other interactions were statistically insignificant ($p > 0.05$).

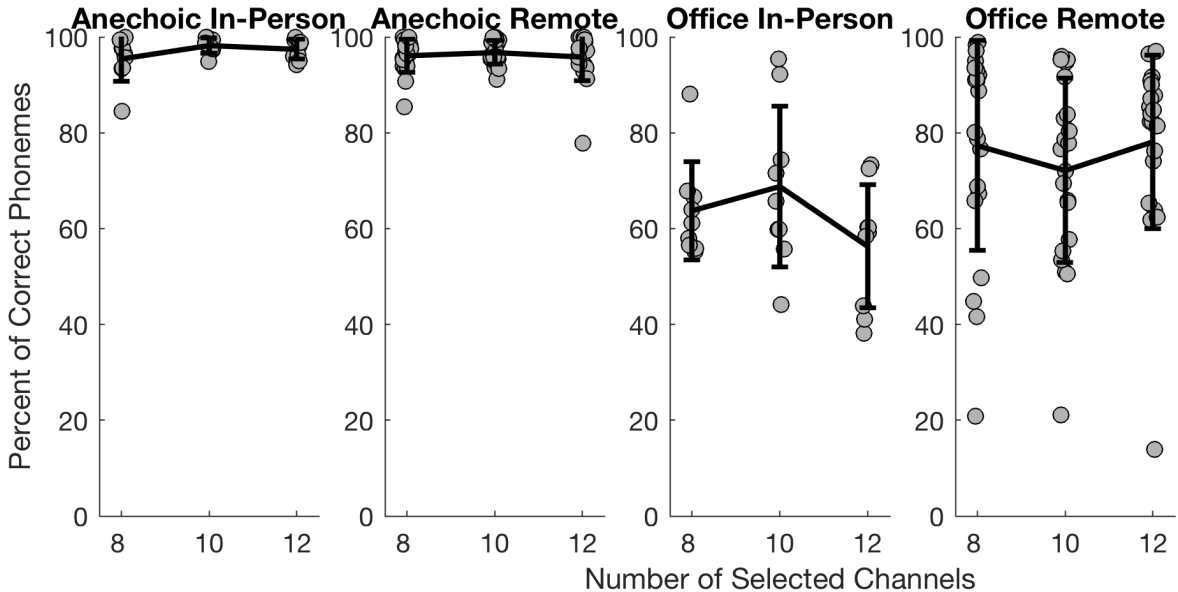

**Fig. 2**. Comparison between speech intelligibility for in-person and remote experiments in an anechoic environment and a reverberant office with 8, 10, or 12 channels selected during each stimulation cycle. The solid line shows the mean percent of correctly identified phonemes, error bars show ± 1 standard deviation, and gray circles indicate individual subject performance.

## 4. Discussion

The goal of this work was to develop and evaluate a framework for conducting speech intelligibility tests remotely. We developed an application that allows participants to run the listening study on their own computers. We compared the results of the remote listening tests with those obtained during in-lab experiments in an anechoic environment and a reverberant office. In the anechoic environment, the mean and standard deviation of the speech intelligibility scores were similar between remote and in-person listening tests. This result suggests that the speech stimuli for the remote listening tests were audible and that there were no major technical issues. In the office, the speech intelligibility was higher and more variable for the remote listening tests than for the in-person tests. The increased intelligibility in the remote study may have resulted from the psychophysical stimulus calibration procedure, as no limit was placed on how high subjects were allowed to set the sound level, whereas the sound level was strictly controlled for the in-person study. Thus, it is possible that subjects from the remote study listened to the sentences at a higher sound level, which could explain the higher speech intelligibility scores obtained in the remote study.

Compared to a traditional laboratory setting, the remote testing framework introduces greater variability due to factors such as differences in sound level, frequency response of headphones, environmental noise, and visual distractions. One technical issue we experienced was a lack of compatibility with wireless Bluetooth headphones, which did not output an audio signal and in some cases caused the testing interface to freeze. This is consistent with findings by the Task Force on Remote Testing, who reported that interference from nearby Bluetooth devices may cause portions of the acoustic signal to be dropped (Stecker et al., 2020). Therefore, the subjects who owned only wireless headphones were instructed to use their computer speakers instead. However, loudspeakers are not desirable in a remote setting because the increased levels of reverberation and noise can detrimentally affect the listener's perception of the acoustic signal (Hughes et al., 2012). This can be mitigated by requiring subjects to use wired headphones, as we did not experience any technical issues with wired headphones. Another limitation of the current setup is that the installer associated with the listening test interface requires an internet connection to download MATLAB Runtime, which in some cases caused the installation process to take nearly one hour due to a slow internet connection. This can be potentially mitigated by deploying the application in a web browser using platforms such as MATLAB Web App Server, which circumvents the need to install programs on the subject's computer.

## 5. Conclusion

In this study, we developed and tested a framework for conducting speech intelligibility tests remotely using a computer program that subjects could run on their own computers. Compared to in-person listening tests, the remote listening tests showed similar levels of performance in an anechoic environment but higher and more variable levels of performance in a reverberant environment. Future work will aim to improve upon the remote testing framework by requiring the use of headphones to reduce the variability in speech intelligibility scores.

**Acknowledgements**

This study was funded by the National Institutes of Health under grant number R01DC014290-05. Support for the Duke Office of Clinical Research to host REDCap was made possible by grant UL1TR001117 from the National Center for Research Resources, a component of the National Institutes of Health (NIH), and NIH Roadmap for Medical Research. The authors would also like to thank the subjects who participated in the in-person and remote studies.

**References and links**